\documentclass[final]{siamltex}

\usepackage{amssymb}
\usepackage{amsfonts}
\usepackage{amsmath}
\usepackage{graphicx}

\def\g{\hat g}

\def\I{\mathcal{I}}

\def\r{\rho}

\def\g{\gamma}

\def\r{\rho}

\def\({\left(}
\def\){\right)}

\def\be#1\ee{\begin{equation}#1\end{equation}}
\def\ba#1#2\ea{\begin{array}#1#2\end{array}}
\def\bgr#1\egr{{\allowdisplaybreaks\begin{gather}#1\end{gather}}}
\def\bma#1\ema{{\allowdisplaybreaks\begin{align}#1\end{align}}}

\def\oplem#1{\begin{lemma}\, {\rm #1}\, \it }
\def\cllem{\end{lemma}\rm \par }
\def\opthm#1{\begin{theorem}\, {\rm #1}\, \it }
\def\clthm{\end{theorem}\rm \par }

\newcommand{\fer}[1]{(\ref{#1})}
\newcommand{\bq}{\begin{equation}}
\newcommand{\eq}{\end{equation}}
\def\bqa{\begin{eqnarray}}
\def\eqa{\end{eqnarray}}
\def\bd{\begin{displaymath}}
\def\ed{\end{displaymath}}
\def\Remark{\medskip\noindent {\bf Remark.}\ \ignorespaces}

\def\gg{\gamma}
\newcommand{\To}{\longrightarrow}
\newcommand{\ind}{\,{1}\!\!\mathbb{I}}

\title{First--order continuous models of opinion formation\thanks{The
authors acknowledge support from 2005 INDAM project ``Kinetic
Innovative Models for the Study of the Behavior of Fluids in
Micro/Nano Electromechanical Systems'', Italian M.I.U.R.\ project
``Mathematical Problems of Kinetic Theories'' and Italian
M.I.U.R.\ project ``Numerical Modelling for Scientific Computing
and Advanced Applications''.}}

\author{
Giacomo Aletti\thanks{Department of Mathematics, University of
Milano, via Saldini 50, 20133 Milano, Italy.
\texttt{giacomo.aletti@unimi.it}}
 \and
 Giovanni Naldi\thanks{Department of Mathematics, University of Milano, via Saldini 50, 20133 Milano,
Italy. \texttt{giovanni.naldi@mat.unimi.it}}
 \and
 Giuseppe Toscani\thanks{Department of
Mathematics, University of Pavia, via Ferrata 1, 27100 Pavia,
Italy. \texttt{giuseppe.toscani@unipv.it}} }

\begin{document}

\maketitle

\begin{abstract}
We study certain nonlinear continuous models of opinion formation derived from
a kinetic description involving exchange of opinion between individual agents.
These models imply that the only possible final opinions are the extremal ones,
and are similar to models of pure drift in magnetization. Both analytical and
numerical methods allow to recover the final distribution of opinion between
the two extremal ones.
\end{abstract}



\pagestyle{myheadings}
\thispagestyle{plain}
\markboth{G. ALETTI, G. NALDI AND G. TOSCANI}{FIRST--ORDER CONTINUOUS MODELS OF OPINION FORMATION}

\section{Introduction}

This paper is devoted to the analysis and large-time behavior of solutions of the equation
 \be\label{magnet}
\frac{\partial f}{\partial t} = \gamma \frac{\partial }{\partial
x}\left((1-x^2)(x - m(t))
 f\right)
 \ee
where the unknown $f(x,t)$ is a time--dependent probability density which
may represent the density of opinion in a community of agents. This opinion varies
between the two extremal opinions represented by $\pm 1$, so that $x \in
\mathcal{I}= [-1,1]$. The constant $\gamma $ is linked to the spreading
($\gamma = -1$) or to the concentration ($\gamma = +1$) of opinions. In
\fer{magnet} $m(t)$ represents the mean value of $f(\cdot, t)$,
 \be\label{mean}
 m(t) = \int_\I f(x,t) \, dx ,
 \ee
 and its presence introduces a nonlinear effect into its evolution.
The value of the constant $\gamma$ is responsible of completely different
effects. In case $\gamma = -1$, this equation, from now on called nonlinear
decision equation, arises in the study of opinion formation, and has been
introduced in \cite{to2}, in connection with the quasi-invariant opinion limit
of a model Boltzmann equation for the kinetic description of opinion formation
involving exchange of opinion between individual agents. The equilibrium state
is given by two Dirac masses located at the extremal points of the interval
$\I$. In case $\gamma = +1$, the equation represents a simple type  of
nonlinear friction equation, and its effect is to concentrate the solution. In
this case, the equilibrium state is given by a Dirac mass located at some point
of the interval $\I$.

When $\gamma = -1$, related equations have been introduced recently. In
\cite{SL} a linear equation describing the pure drift in magnetization has been
derived as the mean field limit of the Sznajd model \cite{SWS} in case of two
opinions. There the (linear) first-order partial differential equation
coincides with equation \fer{magnet}, where it is assumed that $m(t)$ is
identically zero. This simplification allows for an analytical treatment, since
it is possible to obtain the exact solution and control the rate of decay
towards the equilibrium for all values of $\gamma$. If $\gamma =+1$ related
models of one-dimensional nonlinear friction equations have been considered in
the study of granular flows in \cite{McNY2, to1}, in connection with the
quasi-elastic limit of a model Boltzmann equation for rigid spheres with
dissipative collisions and variable coefficient of restitution.

Microscopic models of both social and political phenomena
describing collective behaviors and self--organi\-za\-tion in a
society have been recently introduced and analyzed by several
authors \cite{GGS, Lig, Och, Sta, SO, SWS, Wei, WB}.  The leading
idea is that collective behaviors of a society composed by a
sufficiently large number of individuals (agents) can be hopefully
described using the laws of statistical mechanics as it happens in
a physical system composed of many interacting particles. The
details of the social interactions between agents then
characterize the emerging statistical phenomena.

Among others, the modeling  of opinion formation attracted the interest of a
increasing number of researchers (cfr. \cite{DNAW, Lig, Och, SWS} and the
references therein). Mean fields model equations have been proposed in
\cite{Ben, SL}. These equations are in general partial differential equations
of diffusive type, that can in some case be treated analytically to give
explicit steady states. A kinetic description based on two-body interactions
involving both compromise and diffusion properties in exchanges between
individuals has been proposed in \cite{to2}. Compromise and diffusion were
quantified by two parameters, which are mainly responsible of the behavior of
the model, and allow for a rigorous asymptotic analysis. In a compromise
dominated regime, the resulting equation is our equation \fer{magnet}.

The mathematical methods we use are close to those used in the context of
kinetic theory of nonlinear friction equations \cite{LT}, and made popular by
the mass transportation community \cite{Vil}.

The paper is organized as follows. In the next section we introduce the main
properties of the model, which justify the treatment in terms of a suitable
weak formulation. The qualitative analysis is given in Section 3. The
large--time behavior is considered in Sections 4--6. It is shown that the
problem can be solved in sufficiently high generality only in the case of
concentration ($\g=1$). This lack of generality in the analytical treatment of
the large--time behavior of the solution in the spreading of opinion justifies
the numerical treatment of the equation. The numerical approximation is
included in Section 7.


\section{Main properties and weak description}\label{model}

As briefly described in the introduction, equation \fer{magnet}
describes the evolution of a probability density which represents
the density of opinions in a community. For all values of the
constant $\gamma$, we will show that the time-evolution driven by
this equations leads the density towards a equilibrium state that
is described in terms of two Dirac masses ($\gamma = -1$) or to a
unique Dirac mass ($\gamma =1$). Having in mind that the
equilibrium solution to equation \fer{magnet} is given by Dirac
masses, any convergence result towards equilibrium holds in
weak$^*$-measure sense. The recent analysis of \cite{LT} of the
nonlinear friction equation introduced by McNamara and Young
\cite{McNY2}, suggests that a suitable way of treating equation
\fer{magnet} is based on a rewriting of this equation in terms of
pseudo-inverse functions. It is immediate to show that the drift
operator on the right-hand side of \fer{magnet} preserves
positivity and mass,
\begin{equation}\label{mass}
 \int_\I f(x,t)\, dx = \int_\I f_0(x)\, dx .
\end{equation}
Then, given a initial datum which is a probability density (nonnegative and
with unit mass), the solution remains a probability density at any subsequent
time. Let $F(x)$ denote the probability distribution induced by the density
$f(x)$,
\begin{equation}\label{distr}
 F(x) = \int_{(-\infty,x]} f(y)\, dy
\end{equation}
 and let $\mu$ denote the distribution on $\mathbb R$ associated to $F$.
Since $F(\cdot)$ is not decreasing, we can define  its pseudo
inverse function (also called quantile function) by setting, for
$\rho \in (0,1)$,
 \[
 X^\mu(\rho) = X^F(\rho) = \inf \{ x: F(x) \geq \rho\}
 \]
Equation \fer{magnet} for $f(x,t)$ takes a simple form if written in terms of
its pseudo inverse $X(\rho,t)$. Theorem~\ref{thm:eq-st-we} shows in fact that the
evolution equation for $X(\rho,t)$ reads
 \begin{equation}\label{inv}
\frac{\partial X(\rho,t)}{\partial t}= -\gamma
\left(X(\rho,t)-m(t)\right)(1-X^2(\rho,t)) ,
 \end{equation}
 where now $\rho\in (0,1)$. Note that if we assume $F$ to be
absolutely continuous with respect to $x$ and strictly increasing, then Theorem~\ref{thm:eq-st-we} reduces to elementary computations.
 In \fer{inv}
 \be\label{mean:inv}
 m(t) = \int_0^1 X(\r, t) \, d\r .
 \ee
Let us set $\g = -1$ (spreading). Then, the weak form \fer{inv} clarifies the
evolution of $X(\r, t)$ and the role of $m(t)$. In fact, if $X(\r, t) > m(t)$,
$X(\r ,t)$ increases towards $1$, while $X(\r, t) < m(t)$ implies that $X(\r
,t)$ decreases towards $-1$. Hence, the mean opinion $m(t)$ represent a barrier
for the density of opinions to move towards one of the two extremal opinions.
The fact that the mean opinion varies with time makes the nonlinear problem
harder to handle with respect to the linear problem considered in \cite{SL}
where the barrier is fixed equal to zero.

Among the metrics which can be defined on the space of probability measures,
which metricize the weak convergence of measures \cite{Zol}, one can consider
the $L^p$-distance ($1\leq p <\infty$) of the pseudo inverse functions
\begin{equation}\label{metric}
d_p(X,Y) = \Big( \int_0^1 |X(\r) - Y(\r)|^p \, d\r\Big)^{1/p}.
\end{equation}
In what follows, we'll use the usual identifications $$d_p(X,Y) =
d_p(f_X,f_Y) = d_p(F_X,F_Y) = d_p(\mu_X,\mu_Y)\,, $$ where $\mu_X$
($\mu_Y$), $F_X$ ($F_Y$) and $f_X$ ($f_Y$) denote the
distribution, the cumulative function and the density associated
to $X$ ($Y$), respectively. By this identification, as one can see
\cite{LT,Vas,Vil}, $d_2(F,G)$ is nothing but the Wasserstein
metric \cite{Vas}.

In addition to nonlinear friction equations arising in the modelling of
granular gases \cite{LT}, the strategy of passing to pseudoinverse functions
has been recently applied to nonlinear diffusion equations of porous medium
type \cite{CGT}, and to degenerate convection--diffusion equations \cite{CF}.
This rewriting of nonlinear diffusion equations has been shown to be useful in
order to obtain simple explicit numerical schemes that satisfy a contraction
property with respect to the Wasserstein metric \cite{GoTo}.

\section{Existence, uniqueness and well--position of the problem}
In this section we will study the initial value problem for equation
\fer{magnet}, with initial density
 \be\label{magnet1}
 f(x,t =0)=f_{0}(x), \quad x \in \I.
 \ee
 As before, we will denote by $X_0(\r)$ the quantile function corresponding to
 $f_0$, so that
 \be\label{inv1}
 X(\r,t =0)=X_{0}(\r) = \inf \{ x: F_0(x) \geq \rho\}, \quad \r \in [0,1].
 \ee
 The equivalence between equations \fer{magnet} and \fer{inv} is contained into
 the following.

\begin{theorem}\label{thm:eq-st-we}
There exists a weak solution of \eqref{magnet}--\eqref{magnet1} if and only if
there exists a solution of \eqref{inv}--\eqref{inv1}.
\end{theorem}
\begin{proof}
Suppose first that there exists  $f(x,t)$ which solves
\eqref{magnet}--\eqref{magnet1}. Then $m(t)$ is a differentiable function of
time. Let $y(t)$ be the maximal $C^1$ solution of the Abel
differential equation:
\begin{equation}\label{eq:abel_y}
\left\{%
\begin{array}{l}
y'=-\gamma (1-y^2)(y-m(t))\\
y(0)=\bar y_0 \\
\end{array}%
\right.
\end{equation}
where, for any $y_0\in[-1,1]$ we denoted by $\bar y_0$ a $C^1$--extension of
$y_0$ to $\mathbb R$. We have, in weak sense,
\begin{multline}\label{eq:ev_quant}
\frac{d}{d t} \int_{(-\infty,y(t)]} f(x,t)\,dx = \int_{\mathbb R}
\Big[\frac{\partial}{\partial t} \Big(\ind_{(-\infty,y(t)]}(x)
\Big) f(x,t) +
\ind_{(-\infty,y(t)]}(x) \frac{\partial}{\partial t} f(x,t) \Big]\,dx \\
\begin{aligned}
&= \int_{\mathbb R} \big( y'(t) \delta_{y(t)}(x) +\gamma
[\delta_{y(t)}(x) (1-x^2)(x-m(t)) ]\big) f(x,t)\,dx  \\
&= 0 \,.
\end{aligned}
\end{multline}
Since $X(\rho,t)\leq x \iff \rho \leq \int_{-\infty}^xf(y,t)\,dy$,
the first part of the proof has been shown.

Now, let $X(\rho,t)$ be a solution of \eqref{inv}--\eqref{inv1}. As a
consequence of the properties of the solution to Abel's equation
\eqref{eq:abel_y}, given any initial datum $X(\rho,0)$ satisfying
\begin{itemize}
    \item $X(\rho,0)\in[-1,1]$;
    \item $X(\rho,0)$ is nondecreasing;
    \item $X(\rho,0)$ is left-continuous.
\end{itemize}
the same properties are preserved at any subsequent time $t>0$. Hence, for any
$t$, $\{X(\rho,t),\rho\in(0,1)\}$ is the quantile function of a unique
probability measure on $[-1,1]$. We have only to prove that \eqref{magnet}
holds. This is a consequence of the change of variables formula. In fact,
if $h$ is a test function
\begin{align*}
\int_{\mathbb R} h(x) \frac{\partial}{\partial t}f(x,t)\,dx &=
\frac{\partial}{\partial t} \int_{\mathbb R} h(x) f(x,t)\,dx
\\
& = \frac{\partial}{\partial t} \int_0^1 h(X_\rho(t)) \, d\rho
\\
& = \int_0^1 \frac{\partial}{\partial t} h(X_\rho(t)) \, d\rho
\\
& = \int_0^1 h'(X_\rho(t))
\Big(-\gamma
\left(X(\rho,t)-m(t)\right)(1-X^2(\rho,t))\Big)
\, d\rho
\\
& = \int_{\mathbb R} h'(x)
\big(-\gamma
\left(x-m(t)\right)(1-x^2)\big) f(x,t)\,dx
\\
& = \int_{\mathbb R} h(x)
\frac{\partial}{\partial x} \Big(\gamma
\left(x-m(t)\right)(1-x^2) f(x,t) \Big) \,dx
\,.
\end{align*}
\end{proof}

We call $(\mathcal K,p)$ the (compact) set of probability
distributions on $[-1,1]$ equipped with the $p$-Wasserstein
distance. Note that all the $p$-Wasserstein distances on
$(\mathcal K,d)$ are equivalent. In fact, if $q\geq p\geq 1$,
 \be\label{eq:mon_Lp}
\|X^{\mu_1}-X^{\mu_2}\|_p \leq \|X^{\mu_1}-X^{\mu_2}\|_q \leq
2^{1-p/q}\|X^{\mu_1}-X^{\mu_2}\|_p^{p/q} \,.
 \ee
We will refer to $\mathcal K$ as the topological space of probability
distributions on $[-1,1]$ induced by any of this metric: the
weak$^\ast$-topology. Before searching for a continuous solution of
\eqref{magnet} in $\mathcal K$, we state the following trivial lemma.
\begin{lemma}[Solution Abel]\label{lem:fun}
Let $\phi(x,y)=-\gamma(1-x^2)(x-y)$. Then
$$
|\phi(x_1,y_1)-\phi(x_2,y_2)|\leq 4|x_1-x_2|+|y_1-y_2| \,.
$$
Moreover, if $f$ is a solution of $f'(t)=\phi(f(t),g(t))$ with $\sup|g(t)|\leq
1$ and $f(0)\in[-1,1]$,
$$
|f(s)-f(t)|\leq 2|s-t|\,.
$$
\end{lemma}
We call \emph{solution of \eqref{magnet}--\eqref{magnet1}} any
function $ \mu \in C^0 ( \mathbb R ,\mathcal K)$ s.t.\
\eqref{magnet}--\eqref{magnet1} holds. We have the following
\begin{theorem}\label{th:exist}
For any probability density $f_0(x)$ in \eqref{magnet1}, there
exists a unique function $\mu \in C^0 (\mathbb R ,\mathcal K)$
such that, if  $f(x,t)$ denotes the weak derivative of the
probability distribution $\mu(t)$, $f(x,t)$ satisfies
\eqref{magnet} with initial value \eqref{magnet1}. Moreover, for
any $t\in\mathbb R$, the solution depends in a continuous way on
the initial datum: the problem \eqref{magnet}--\eqref{magnet1} is
well--posed in $C^0 (\mathbb R ,\mathcal K)$.
\end{theorem}
\begin{proof}[Existence]
We prove the existence of a solution of the equivalent problem
\eqref{inv}--\eqref{inv1} (see Theorem~\ref{thm:eq-st-we}) in a
constructive way. More precisely,
\begin{description}
    \item[A] we construct a sequence $\{X_n,n\in\mathbb N\}$
    which approximates a target solution;
    \item[B] by compactness arguments, we find a
    convergent subsequence $X_{n_l}\to X$;
    \item[C] the limit $X$ satisfies \eqref{inv}--\eqref{inv1}.
\end{description}

\medskip
Let $[-T,T]$ be fixed. For any $n\in\mathbb N$, we subdivide
$[-T,T]$ into disjoint intervals of length $R/2^N$. Then we
proceed as follows:
\begin{description}
    \item[A1] we compute $m^{(n)}_0=\int_{0}^1
    X_0(\rho)\,d\rho$;
    \item[A2] we solve \eqref{inv} on $[-T/2^n,T/2^n]$ with
    $m^{(n)}(t)=m_0$, finding $X^{(n)}(\rho,t)$,
    $ t\in [-T/2^n,T/2^n]$;
    \item[A3] for any $k=1,\ldots, 2^n-1$,
    \begin{itemize}
        \item we compute
        \[
        m^{(n)}_{\pm k}= \int_{0}^1 X^{(n)}(\rho,\pm kT/2^n)\,d\rho ;
        \]
        \item we solve \eqref{inv} on $(kT/2^n,(k+1)T/2^n] $
        with $m^{(n)}(t)=m^{(n)}_{k}$ and initial data $X^{(n)}(\rho,kT/2^n)
        $, finding $X^{(n)}(\rho,t)$,
        $ t\in(kT/2^n,(k+1)T/2^n] $;
        \item we solve \eqref{inv} on $[-(k+1)T/2^n,-kT/2^n) $
        with $m^{(n)}(t)=m^{(n)}_{-k}$ and initial data $X^{(n)}(\rho,-kT/2^n)
        $, finding $X^{(n)}(\rho,t)$,
        $ t\in[-(k+1)T/2^n,-kT/2^n) $.
    \end{itemize}
\end{description}
We call $\mu^{(n)}:[-T,T]\to\mathcal K$ the sequence of function
with value in $\mathcal K$ associated to $X^{(n)}$.

\noindent\textbf{B}\phantom{.} For any $n\in\mathbb N$, it is
possible to prove (by induction on $k$) that for any $t\in[-T,T]$
$X^{(n)}(\rho,t)\in[-1,1]$ and $m^{(n)}(t)\in[-1,1]$. As a
consequence of Lemma~\ref{lem:fun}, we have
\begin{equation}\label{eq:lem_0}
\big|X^{(n)}(\rho,s)-X^{(n)}(\rho,t)\big| \leq 2 |t-s|
\end{equation}
i.e., for any $\rho\in(0,1)$,
$\{X^{(n)}(\rho,\cdot):[-T,T]\to[-1,1]\}_{n\in\mathbb N}$ is a
uniformly equicontinuous sequence. A diagonal argument together
with Ascoli-Arzel\`a Theorem ensure the existence of a subsequence
$n_l$ s.t.\
$\{X^{(n_l)}(\rho,\cdot):[-T,T]\to[-1,1]\}_{l\in\mathbb N}$
converges uniformly on $[-T,T]$ for each $\rho\in(0,1)\cap\mathbb
Q$.

Now, \eqref{eq:lem_0} implies
\begin{equation}\label{eq:lem_1} \int_{0}^1 |X^{(n)}(\rho,s)-X^{(n)}(\rho,t)|\,
d\rho \leq 2 | t- s | \,,
\end{equation}
i.e., $\mu^{(n)}$ is a equicontinuous sequence with respect to the
distance $d_1$ on $\mathcal K$. Then, Ascoli-Arzel\`a Theorem
again ensures the existence of a subsection of $n_l$ (we call it
$n_l$ again) s.t.\
\begin{align}\label{eq:AA1}
&\sup_{t\in[-T,T]} d_1 ( \mu^{(n_k)}(t),\mu^{(n_l)}(t)) \leq
M(k\wedge l) \mathop{\To}\limits_{k\wedge l\to\infty} 0
\\
\label{eq:AA2} & \sup_{t\in[-T,T]} |X^{(n_k)}(\rho,t) -
X^{(n_l)}(\rho,t) | \leq N_\rho(k\wedge l)
\mathop{\To}\limits_{k\wedge l\to\infty} 0 \,,&&\forall
\rho\in\mathbb Q\cap(0,1)
\end{align}
Now, let $\rho\in(0,1)\cap\mathbb Q$ be fixed.
$\{X^{(n_l)}(\rho,\cdot)\}_{l\in\mathbb N}$ is a uniform
convergent sequence of derivable functions converging to
$X(\rho,t)=\lim_l X^{(n_l)}(\rho,t)$). Left-continuity and
monotonicity of $\{X(\rho,t),\rho\in(0,1)\cap\mathbb Q\}$ extend
the definition of $X(\rho,t)$ to all $\rho\in(0,1)$.

\noindent\textbf{C}\phantom{.} What remains to prove is
\begin{itemize}
    \item\label{pr:1} $\lim_l m^{(n_l)} (t) = \int_0^1
    X(\rho,t)\,d\rho=:m(t)$;
    \item\label{pr:2} $X(\rho,t)$ is differentiable, and
    \eqref{inv} holds.
\end{itemize}
Note that, from \eqref{eq:AA1}, it follows immediately that $ |
\int_0^1 X^{(n_k)}(\rho ,t) \,d\rho- \int_0^1 X^{(n_l)}(\rho ,t)
\,d\rho |$ $ \leq M(k\wedge h) $ and hence
\begin{equation}\label{eq:AAm11}
\Big| \int_0^1 X(\rho,t) \,d\rho- \int_0^1 X^{(n_l)}(\rho,t)
\,d\rho \Big| \leq M(h)\,.
\end{equation}
By definition of $m^{(n)}$,
$$
m^{(n)}(t)=m^{(n)}([[2^nt]]/2^n)= \int_{0}^1
X^{(n)}(\rho,[[2^nt]]/2^n) \, d\rho
$$
where $[[\cdot]]$ is the integer part of $\cdot$ closer to $0$.
Therefore,
\begin{equation}\label{eq:inc_mean1}
\Big|m^{(n)}(t)-\int_{0}^1 X^{(n)}(\rho,t) \, d\rho \Big| \leq
2^{-n+1}
\end{equation}
and hence $m^{(n_l)}(t)$ is a Cauchy sequence on $[-1,1]$. Thus,
there exists $\widehat{m}(t)=\lim_h m^{(n_l)}(t)$. By
\eqref{eq:AAm11} and \eqref{eq:inc_mean1} it follows that
$\widehat{m}(t)=m(t)$ since
\begin{equation*}
\Big|\widehat{m}(t)- \int_0^1 X(\rho,t)\, d\rho \Big| \leq
\big|\widehat{m}(t)- m^{(n_l)}(t)\big|+ 2^{-n_l+1} +M(h)\,.
\end{equation*}

Now, let $\rho\in(0,1)\cap\mathbb Q$ be fixed. For simplicity of
notation, define
 $\displaystyle
H(n,\rho,t):=\left.\frac{\partial}{\partial s} X^{(n)}(\rho
,s)\right|_{s=t} $. Moreover, we define
 \[
 H(\rho,t) := \lim_{l\to\infty} H(n_l,\rho,t)
=-\gamma(1-X(\rho,t)^2)(X(\rho,t) -m(t)).
 \]
 The uniform convergence theorem
states that $\displaystyle \left.\frac{\partial}{\partial s}
X(\rho,s)\right|_{s=t} = H(\rho,t) $ if $\{H(n_l,\rho,t)
\}_{l\in\mathbb N}$ is a uniform converging sequence on $(-T,T)$.
To prove this, let $k\geq l$. Triangular inequality
\begin{equation*}
\begin{aligned}
\big|H({n_k},\rho,t)-H({n_l},\rho,t)\big| &\leq
\big|H({n_k},\rho,t)-H({n_k},\rho,[[2^{n_l}t]]/2^{n_l})\big| \\
&\qquad +
\big|H({n_k},\rho,[[2^{n_l}t]]/2^{n_l})-H({n_l},\rho,[[2^{n_l}t]]/2^{n_l})\big|
\\
&\qquad\qquad +
\big|H({n_l},\rho,[[2^{n_l}t]]/2^{n_l})-H({n_l},\rho,t)\big|
\\
& = A_\rho(k,l,t) + B_\rho(l,k,t) + A_\rho(l,l,t)
\end{aligned}
\end{equation*}
shows that we may prove that $ \sup_{t\in[-T,T]} A_\rho(l,k,t) +
B_\rho(l,k,t) + A_\rho(l,l,t) \mathop{\To}\limits_{l\wedge
k\to\infty} 0 $. As a consequence of \eqref{eq:lem_1} and
\eqref{eq:inc_mean1}, we have $ |m^{(n)}(s)-m^{(n)}(t)| \leq
2^{-n+2} +2|t-s|$, which implies (see Lemma~\ref{lem:fun} and
\eqref{eq:lem_0})
\begin{equation*}
\begin{aligned}
|H(n,\rho,s)-H(n,\rho,t)| &= \Big|
(1-X^{(n)}(\rho,s)^2)(X^{(n)}(\rho,s) -m^{(n)}(s))
\\
& \qquad - (1-X^{(n)}(\rho,t)^2)(X^{(n)}(\rho,t) -m^{(n)}(t))
\Big|
\\
&\leq 4 | X^{(n)}(\rho,s)- X^{(n)}(\rho,t)| +
|m^{(n)}(s)-m^{(n)}(t)|
\\
& \leq 10 |t-s|+2^{-n+2}
\end{aligned}
\end{equation*}
and hence $ A_\rho(k,l,t) \leq 10 \cdot 2^{-{n_l}} +2^{-{n_k}+2}
$. Now, let $k\geq l$ and $l\in\{-2^{n_l}+1,\ldots,2^{n_l}-1\}$.
Again, Lemma~\ref{lem:fun}, \eqref{eq:AA1} and \eqref{eq:AA2}
imply $$\big|H({n_k},\rho,l/2^{n_l})-H({n_l},\rho,l/2^{n_l})\big|
\leq 4 N_\rho(l) + M(l) $$ and hence $ B_\rho(k,l,t) \leq
4N_\rho(l)+M(l)
 $. This completes the proof for $\rho\in(0,1)\cap\mathbb Q$.
Now, fixed $y_0 \in[-1,1]$, let $y(t)$ be the maximal $C^1$
solution of the Abel differential equation \eqref{eq:abel_y}.
Since $ X(\rho,t) \geq y(t) \iff X(\rho,0) \geq y(0) $, $\forall
\rho\in(0,1)\cap\mathbb Q $, left-continuity and monotonicity of
$\{X_\rho(t),\rho\in(0,1)\}$ extend the proof to $\rho\in(0,1)$.

\bigskip\noindent
[Uniqueness and well-position] Let $Y(\rho,t), X(\rho,t)$ be two
solutions of \eqref{magnet}--\eqref{mean}. Denote by $m_X(t)$ and
$m_Y(t)$ the mean values of $X$ and $Y$ respectively at time $t$.
Since $ |m_Y(t)- m_X(t) | \leq d_1(X,Y) $ we have (by
Lemma~\ref{lem:fun} and \eqref{eq:mon_Lp})
\begin{multline}\label{eq:Gronw}
\frac{d}{dt} d_2(\mu_X(t),\mu_Y(t)) =
\frac{d}{dt}\int_0^1 \big(Y(\rho,t)- X(\rho,t)\big)^2\,d\rho \\
\begin{aligned}
& \leq 2\int_0^1 \big|Y(\rho,t)- X(\rho,t) \big| \Big(
4\big|Y(\rho,t)- X(\rho,t) \big| + \big|m_Y(t)- m_X(t) \big|
\Big)\,d\rho
\\
& \leq 10\cdot d_2(\mu_X(t),\mu_Y(t)) \,.
\end{aligned}
\end{multline}
Gronwall's lemma completes the proof.
\end{proof}

\section{Large--time behavior of solutions}
Thanks to the uniqueness of solutions of Abel equation, we obtain the following
\begin{lemma}\label{lem:massification}
For any $t \not= s\in\mathbb R$ and $\rho \not= \rho'\in (0,1)$, we have
$$
X(\rho,t)=X(\rho',t) \iff X(\rho,s)=X(\rho',s)\,,
$$
i.e.\ \eqref{magnet} does not create or destroy delta masses in finite time.
\end{lemma}

A direct consequence of the previous lemma is that the initial masses in $+ 1$,
$-1$ and in $(-1,1)$ remain unchanged in time. Let us call them $p_{+1}$,
$p_{-1}$ and $1-(p_{+1}+p_{-1})$, respectively.

An important argument linked to the large--time behavior of solutions to
nonlinear equations is the study both of conservation laws and of Lyapunov
functionals. In addition to mass conservation, a second conserved quantity
(when defined) is furnished by
 \be\label{cons1}
 T(t):= \int_0^1 \log\Big(\frac{1+X(\rho,t)}{1-X(\rho,t)} \Big)
\,d\rho\,.
 \ee
 In addition to the conservation of both mass and $T(t)$, equation \fer{magnet}
 possesses a Lyapunov functional, simply given by the variance of the solution
 \be\label{var}
V(t):= \int_0^1 (X(\rho,t))^2 \,d\rho - \Big(\int_0^1 X(\rho,t) \,d\rho
\Big)^2\,.
 \ee
We give below an easy to check condition which insure both the the boundedness
and the conservation in time of the functional \fer{cons1}.

\begin{lemma}\label{lem:FUNZ_TOSC}
Let $\log\big((1+X(\rho,0))/(1-X(\rho,0))\big)\in L^1(0,1)$. Then, for all $t\in\mathbb R$
$$
T(t):= \int_0^1 \log\Big( \frac{1+X(\rho,t)}{1-X(\rho,t)} \Big)
\,d\rho
$$
is well-defined.  Moreover, $T(t)$ is differentiable and $T'(t)=0$.
\end{lemma}
\begin{proof}
Since $\log\big((1+X(\rho,0))/(1-X(\rho,0))\big)\in L^1(0,1)$, then
$X(\rho,0)\in(-1,1)$, $\forall\rho\in(0,1)$.
Now
$
(1-x^2)(x -1) \leq (1-x^2)(x -m(t)) \leq (1-x^2)(x -1)
$,
and hence the uniqueness of solutions of Abel equations imply
$$
X(\rho,t)\in(-1,1)\,, \qquad \forall\rho\in(0,1)\,,\forall t\in\mathbb R \,.
$$
Let $G(\rho,t)= \log\big(
(1+X(\rho,t))/(1-X(\rho,t)) \big)$.
Since $|G_t(\rho,t)| \leq 2$, we have
\begin{itemize}
\item $\exists G_t$, $\forall \rho\in(-1,1),\forall t\in[-T,T]$;
\item $|G(\rho,t)|\leq |G(\rho,0)|+2T$ and hence $T(t)$ exists;
\item $|G_t(\rho,t)|\leq 2$ and $2\in L^1(0,1)$;
\end{itemize}
and hence it is possible to differentiate under the integral
sign, obtaining $T'(t)=0$.
\end{proof}

The following lemma shows that the variance is a Lyapunov
functional for equation \fer{magnet}.

\begin{lemma}\label{lem:VAR}
The variance $V(t)$ of $\mu(t)$ is a monotone differentiable
function with values in $[0, 1]$.
\end{lemma}
\begin{proof}
$V(t)$ is clearly differentiable, and
$$
V'(t) = \frac{d}{dt} \int_0^1 (X_t(\rho)-m(t))^2 \,d\rho= -2\gamma
\int_0^1 (1-X_t(\rho)^2)(X_t(\rho)-m(t))^2 \,d\rho \,.
$$
In case $\g= -1$, $V(t)$ is monotonically increasing, while bounded from above
by 1. In fact, the maximum value of $V$ is attained for
$\mu=(\delta_1+\delta_{-1})/2$. If on the contrary $\g= 1$, $V(t)$ is
monotonically decreasing, while bounded from below by 0. In this second case,
the minimum value of $V$ is attained for $\mu=\delta_a,a\in[-1,1]$.
\end{proof}

The most difficult problem linked to equation \fer{magnet} is the study of the
evolution of the mean $m(t)$, and to the exact evaluation of its limit value
$\bar m = \lim_{t \to \infty} m(t)$. The knowledge of $\bar m$ is of primary
importance, since in consequence of the structure of the limit state of the
solution to equation \fer{inv},  this value is enough to characterize
completely the steady state.

\Remark The difficulty comes out from the evolution of the
mean $m(t)$, which is given by the ``non closed'' equation
 \be\label{mean-ev}
 m'(t) = -\g m(t) \int_0^1 X^2(\r) \, dr + \g \int_0^1 X^3(\r) \, dr.
 \ee

In what follows, we make use of the previous results to extract
information on the behavior of the mean $m(t)$.

\begin{lemma}\label{lem:bound_of_mt}
For any $t$, $m(t)\in [-\sqrt{1-V(t)},\sqrt{1-V(t)}]$.
\end{lemma}
\begin{proof}
Since $V(t)=\int x^2 f (x,t)dx-(m(t))^2$, then $V(t)\leq
1-(m(t))^2$.
\end{proof}

\begin{lemma}\label{mt_in_L2}
The function $m':\mathbb R\to [-1,1]$ belongs to $L^2(\mathbb R)$. Moreover,
$\exists\lim_{t\to\infty} m'(t)=0$.
\end{lemma}
\begin{proof}
It is sufficient to note that
\begin{equation*}
(m'(t))^2 \leq \int \big(1-X_\rho (t)^2\big)^2(X_\rho (t) -m(t))^2
\,d\rho \leq V'(t) \,.
\end{equation*}
and hence $m'\in L^2(\mathbb R)$ by Lemma~\ref{lem:VAR}. Moreover, since $m'$ is a
Lipschitz function
(in fact it is differentiable and $m''\leq 10$, cfr.\ \eqref{mean-ev}), it follows that
$\lim_{t\to\infty} m'(t)=0$.
\end{proof}

\section{Spreading of opinions}

In this section we study the large behavior of $\eqref{magnet}$ when $\g=-1$,
leaving the opposite case $\g= 1$ to the following section.

\Remark If $X(\rho,0)=-X(1-\rho,0)$ (i.e., the initial distribution is
symmetric), then from \fer{mean-ev} follows that $m(t)=0$ for any subsequent
time $t>0$. If, in addition, $X(\rho_1,0)=X(\rho_1+\delta,0)=0$, then
$X(\rho_1,t)=X(\rho_1+\delta,t)=0$ for all $t>0$ (i.e., any initial mass in $0$
is not moved away in time if the initial distribution is symmetric). In order
to avoid these situations we will allow delta masses only in $\pm 1$:
$X(\rho_1,0)=X(\rho_2,0) \iff \rho_1=\rho_2$ or $(X(\rho_1,0))^2= 1$.

\begin{theorem}\label{lem:M_infty}
Assume $X(\rho_1,0)=X(\rho_2,0) \iff \rho_1=\rho_2$ or $(X(\rho_1,0))^2= 1$.
Then,  the limit distribution  exists and it is given by two masses located in
$-1$ and $+1$.
\end{theorem}
\begin{proof}
By Lemma~\ref{lem:VAR} and Lemma~\ref{lem:bound_of_mt},
$m(t)\in[-\sqrt{1-V(0)},\sqrt{1-V(0)}]$, for all $t\ge0$. Thus, if
$(X(\rho,t_0)^2>1-V(0)$, then $(X(\rho,t)^2>1-V(0)$, for all $t\geq t_0$.
Since $ \displaystyle X(\rho,t) < -\sqrt{1-V(0)}\Rightarrow
\frac{\partial}{\partial t}X(\rho,t)\leq 0$ and $\displaystyle X(\rho,t) >
\sqrt{1-V(0)} \Rightarrow \frac{\partial}{\partial t}X(\rho,t)\geq 0$
 by Lemma~\ref{lem:VAR} and Lemma~\ref{lem:bound_of_mt}, the two functions
\begin{align*}
p^{h}_{-1}(t) &= \sup\{\rho\in (0,1)\colon X(\rho,t) <
-\sqrt{1-V(0)} \} = \int_{[-1,-\sqrt{1-V(0)})} f(x,t)\, dx
\\
p^{h}_{1}(t) &= \inf\{\rho\in (0,1)\colon X(\rho,t) >
\sqrt{1-V(0)} \} = \int_{(\sqrt{1-V(0)},1]} f(x,t)\, dx
\end{align*}
are
monotone. We call $ p^{h}_{\pm 1}=\lim_{t\to\infty}p^{h}_{\pm 1}(t)$.
\eqref{inv} and monotonicity of Abel's solutions allows to state that
\begin{align} \label{eq:monot_lim_X1}
&\forall \rho\in [0,p^{h}_{-1}), \lim_{t\to\infty} X(\rho,t) = -1
\\ \label{eq:monot_lim_X2}
&\forall \rho\in (p^{h}_{+1},1], \lim_{t\to\infty} X(\rho,t) = 1 .
\end{align}
Hence the limit distribution has two masses in $-1$ and $+1$. It remains to
characterize what happens for the remaining $p^{h}_{+1}-p^{h}_{-1}$ mass. Let
us recall that
 \begin{equation}\label{eq:contr_01}
X(\rho,t)\in \big[-\sqrt{1-V(0)},\sqrt{1-V(0)}\big]\,, \qquad
\forall \rho\in (p^{h}_{-1},p^{h}_{+1}) \,,
 \end{equation}
By Lemma~\ref{mt_in_L2}, there exists $ T>0$ such that $ |m'(t)|\leq
(V(0)/2)^2$ for all $t>T$. By contradiction, suppose that there exists $t_0
\geq T$ and $ \rho\in (p^{h}_{-1},p^{h}_{1})$ such that
$\big|X(\rho,t_0)-m(t_0)\big|
> V(0)/4$.

Since $\displaystyle |\frac{\partial}{\partial t} X(\rho,t_0)|>|m'(t_0)|$, it
follows that $\big|X(\rho,t)-m(t)\big|
> V(0)/4$, for all $t\geq t_0$. Thus, \eqref{eq:contr_01} shows the contradiction:
\begin{itemize}
    \item $\frac{\partial}{\partial t} X(\rho,t)$ is continuous;
    \item $|\frac{\partial}{\partial t} X(\rho,t)|> (V(0)/2)^2 $
    \quad {\rm if}
     $t\geq t_0$;
    \item $X(\rho,t)$ is bounded.
\end{itemize}
Therefore,
\begin{equation}\label{eq:contr_0}
\big|X(\rho,t)-m(t)\big| \leq \frac{V(0)}{4} \,, \qquad {\rm for\,\, all} \quad
\rho\in (p^{h}_{-1},p^{h}_{+1}), \quad t>T\,.
\end{equation}
Now, let $F(x,y)=(1-x^2)(x-y)$ as in Lemma~\ref{lem:fun}. Since $F$ is
differentiable, when $x_1\geq x_2$, Lagrange theorem states that we can find
$\xi\in(x_1,x_2)$ such that
 \[
F(x_1,y)-F(x_2,y) = (x_1-x_2) \frac{\partial}{\partial x}F(x,y)\Big|_{x=\xi}.
 \]
 Now, if $1-x^2 \geq V(0)$ and $|x-y|\leq V(0)/4$, we have
  \[
\frac{\partial}{\partial x}F(x,y) = 1 - 3x^2 +2xy \geq V(0) + 2x(y-x) \geq
\frac{V(0)}{2},
 \]
 that is
\begin{equation}\label{eq:contr_1}
F(x_1,y)-F(x_2,y) \geq (x_1-x_2)\frac{V(0)}{2}\, , \qquad   x_i^2 \leq 1-V(0)
\text{ and } |x_i-y|\leq \frac{V(0)}{4}\,.
\end{equation}
Let $p^{h}_{-1} < \rho_2\leq\rho_1 < p^{h}_{+1}$. Then, both
\eqref{eq:contr_01} and \eqref{eq:contr_0} are satisfied for all $t>T$,
\eqref{eq:contr_1} holds, and
$$
\frac{\partial}{\partial t} \Big( X(\rho_1,t) - X(\rho_2,t) \Big)
\geq \big( X(\rho_1,t) - X(\rho_2,t) \big) \frac{V(0)}{2} \,,
\qquad \forall t>T \,.
$$
Since the two solutions are bounded, the only possibility is that $ X(\rho_1,t)
= X(\rho_2,t) $, for all $t>T$ , and for all $(\rho_1,\rho_2)\colon p^{h}_{-1}
< \rho_2\leq\rho_1 < p^{h}_{+1}$, which implies $ p^{h}_{-1} = p^{h}_{1} $ by
Lemma~\ref{lem:massification} and hypothesis.
\end{proof}

\Remark Theorem~\ref{lem:M_infty} may be read in terms of weak$^*$-measure
convergence
$$
f(x,t)\mathop{\rightharpoonup}\limits_{t\to\infty}
p^h_{-1}\delta_{-1}(x)+ (1-p^h_{-1})\delta_{1}(x) \,.
$$
In particular, since the support is compact, all the moments exists and
will converge. We have the
following
\begin{corollary}\label{lem:M_infty_m}
Assume $X(\rho_1,0)=X(\rho_2,0) \iff \rho_1=\rho_2$ or
$(X(\rho_1,0))^2= 1$. Then there exists $\lim_{t\to+\infty} m(t)=
m_\infty=1-2p^h_{-1}$.
\end{corollary}

\section{Concentration of opinions}

Let us recall that the stochastic partial order is naturally given on $\mathcal
K$.  Let $F(x),G(x)$ denote two probability distributions and $X_F,X_G$ their
pseudo-inverse functions, respectively. We say that $ F\preceq G$ if $F(x)\geq
G(x),\forall (x)$ or, equivalently, if $X_F(\rho)\leq X_G(\rho),\forall
\rho\in(0,1)$.
\begin{lemma}\label{lem:monoton}
The operator
 \[
 \phi(X) = -(X- m(X))(1-X^2) \, ,
 \]
is a monotone operator with respect the stochastic ordering.
\end{lemma}
\begin{proof}
Assume $X_1(\rho,s)\leq X_2(\rho,s)$, for all $\rho\in(0,1)$. Then
$m_{1}(s)\leq m_{2}(s)$ (they are equal iff the distributions coincide). Let
$\rho\in(0,1)$ be fixed. If $X_1(\rho,s)= X_2(\rho,s)$, then $X'_1(\rho,s)\leq
X'_2(\rho,s)$. The continuity of $ X' $ is sufficient for the remaining part of
the proof.
\end{proof}

\begin{lemma}\label{lem:concm} Let $X_0$ in \fer{inv1} be given. Then, there exists
$\lim_{t\to+\infty} m(t)= m_\infty$.
\end{lemma}
\begin{proof}
Let $[a,b]$ be the class limit of $m(t)$. Suppose $a=-1$, i.e.\
$\liminf_t m(t)=-1$. Markov inequality then shows that the limit
distribution is a mass concentrated in $-1$, and hence $b=-1$.
Otherwise, we may assume that $ m(t) \in [-1+\delta,1-\delta]$,
for all $t\geq t_0$ and let $p_0$ the mass not concentrated in
$\pm 1$ at each time (recall Lemma~\ref{lem:massification}). For
all $\epsilon>0$, $p_0-\epsilon$ mass is in
$[-1+\epsilon,1-\epsilon]$ at $t=t_0$. Therefore for all
$\rho\in(0,1)$ $X(\rho,t_0)\in [-1+\epsilon,1-\epsilon]$, and
$X(\rho,t)$ decades exponentially to $m(t)$ with rate not less
than $(\min(\delta,\epsilon))^2$. The large behavior of this
process shows three delta masses: the initials in $\pm 1$ and the
remaining in $m(t)$. Stationary arguments imply the existence of
$m_\infty$.
\end{proof}

The steady state of the process can now be defined by the following

\begin{theorem}
If $(1-p_{1})(1-p_{-1})<1$ (i.e., if there are masses in $\pm 1$ at time $t=0$)
then $ m_\infty = p_{1}-p_{-1}$. Otherwise, if $\log\big((1+X(\rho,0))/(1-X(\rho,0))\big)\in L^1(0,1)$
then
 \be\displaystyle\label{eq:m_infty}
 m_\infty = \frac{ \exp \left\{T(0)\right\} -1 }{ \exp\left\{T(0)\right\} +1}
 \ee
\end{theorem}
\begin{proof}
The first part is a consequence of Lemma~\ref{lem:concm} and
stationary arguments. The second part is a consequence of
Lemma~\ref{lem:FUNZ_TOSC}, since
$$
\int_0^1 \log \Big( \frac{1+X(\rho,0)}{1-X(\rho,0)} \Big) \,d\rho
= \int_0^1 \log \Big( \frac{1+X(\rho,t)}{1-X(\rho,t)} \Big)\,d\rho
\mathop{\longrightarrow}\limits_{t\to\infty} \log \Big(
\frac{1+m_\infty}{1-m_\infty} \Big) \,,
$$
the last limit being true by Lemma~\ref{lem:concm}.
\end{proof}

\Remark Lemma~\ref{lem:monoton} allows to extend the previous
result to the cases where at least one of the two functions
$\log(1\pm X(\rho,0))$ is integrable. If, for example, $\log(1+
X(\rho,0))\in L^1(0,1)$ and $\log(1- X(\rho,0))\not\in L^1(0,1)$,
if we take $X^{(n)}(\rho,0)=\min\{X(\rho,0),$ $1-1/n\}$, then
$X^{(n)}(\rho,t)\leq X(\rho,t)$, $\forall t\geq 0$, $\forall
\rho\in(0,1)$. Monotone Convergence Theorem states $\lim_n
T^{(n)}(0)=+\infty$, i.e.\ $\lim_n m^{(n)}_\infty= 1$. Thus, by
monotonicity argument of Lemma~\ref{lem:monoton}, $m_\infty= 1$.

With this remark in mind, we show now a ``counterintuitive''
example. Let
$$
f_0 (x)=
 \begin{cases}
\frac{1-\epsilon}{\epsilon} & \text{if }-1< x < -1+\epsilon
 \\
\frac{1}{1-x}\big(\frac {\epsilon}{ 1-\epsilon\log( 1-x)}\big)^2
 & \text{if }0< x < 1
 \end{cases}
$$
and hence
$$
F_0 (x)=
 \begin{cases}
 0 & \text{if }x< -1
 \\
\frac{1-\epsilon}{\epsilon}(1+x) & \text{if }-1\leq x <
-1+\epsilon
 \\
1-\epsilon & \text{if } -1+\epsilon \leq x < 0
 \\
1-\frac{\epsilon}{1-\epsilon\log(1-x)} & \text{if }0\leq x < 1
 \\
1 & \text{if } 1 \leq x
 \end{cases}
$$
which corresponds to
$$
 X_0 (\rho)=
 \begin{cases}
 -1 + \frac{\epsilon}{1-\epsilon}\rho & \text{if }0<\rho\leq 1-\epsilon
 \\
 1-\exp(-\frac{1}{1-\rho}+\frac{1}{\epsilon})  & \text{if
 }1-\epsilon<\rho<1
 \end{cases}
$$
With this data, $\log(1+ X(\rho,0))\in L^1(0,1)$ but $\log(1-
X(\rho,0))\not\in L^1(0,1)$; $1-\epsilon$ initial mass is close to
$-1$ while the asymptotic solution is a $\delta_1$.

\section{Numerical examples}

The analysis of the previous section left open the problem of the
identification of the steady state in the case of the magnetization. Here
results can be achieved only by numerical simulation of the spreading process.
To test the numerical method, we will first derive the (explicit) solution to
the pure drift linear equation of spreading considered in \cite{SL} as the mean
field limit of the Sznajd model \cite{SWS}. This equation reads
 \be\label{magnet-lin}
 \frac{\partial f}{\partial t} = \gamma \frac{\partial
}{\partial x}\left(x(1-x^2)
 f\right),
 \ee
 namely equation \fer{magnet} without the presence of the mean $m(t)$. In terms
 of the quantile function $X(\r,t)$ equation \fer{magnet-lin} takes the form
\begin{equation}\label{inv-lin}
\frac{\partial X(\rho,t)}{\partial t}= -\gamma X(\rho,t)(1-X^2(\rho,t)).
 \end{equation}
Let us set $\g =-1$ (spreading), and let $X_0(\r)$ denote the initial datum.
For any given $\r \in [0,1]$, equation
 \fer{inv-lin} is an ordinary differential equation which can be easily
 integrated to give
 \be\label{sol1}
 X(\r,t) = \frac{X_0(\r) e^t}{\left(1-X_0^2(\r) + X_0^2(\r)e^{2t}
 \right)^{1/2}}.
 \ee
The asymptotic behavior of equation \fer{inv-lin} can be easily
deduced from the explicit solution. In fact, the solution
converges exponentially in time to $-1$ if $X_0(\r) <0$, while it
converges to $+1$ if $X_0(\r)
>0$. Solution \fer{sol1} can be inverted by using the
definition of $X(\r,t)$. Let $F_0(x)$ $x \in \I$ be the initial
distribution function, then, since $X_0(F_0(x)) = x$
 \be\label{sol2}
 X(F_0(x),t) = \frac{x e^t}{\left(1-x^2 + x^2e^{2t}
 \right)^{1/2}}.
 \ee
Thus, since the function on the right of \fer{sol2} is increasing
with respect to the variable $x$, we
 can invert it to obtain
 \be\label{sol3}
 X\left(F_0\left( \frac{y}{\left((1-y^2)e^{2t} +y^2
 \right)^{1/2}}  \right),t\right)
  = y.
 \ee
 Finally, equation \fer{sol3} finally implies
 \be\label{sol4}
 F(y,t) = F_0 \left(\frac{y}{\left((1-y^2)e^{2t} +y^2
 \right)^{1/2}}  \right).
 \ee
 Differentiating with respect to $y$ we conclude that, if $f_0(x)$ $x \in \I$ is
 an initial density for equation \fer{magnet-lin}, the solution in time is given by
 \be\label{sol5}
 f(x,t) =  \frac{e^{2t}}{\left((1-x^2)e^{2t} +x^2
 \right)^{3/2}}f_0 \left(\frac{x}{\left((1-x^2)e^{2t} +x^2
 \right)^{1/2}}  \right).
 \ee

\begin{figure}[ht]
    \centering
    \includegraphics[width=.48\textwidth]{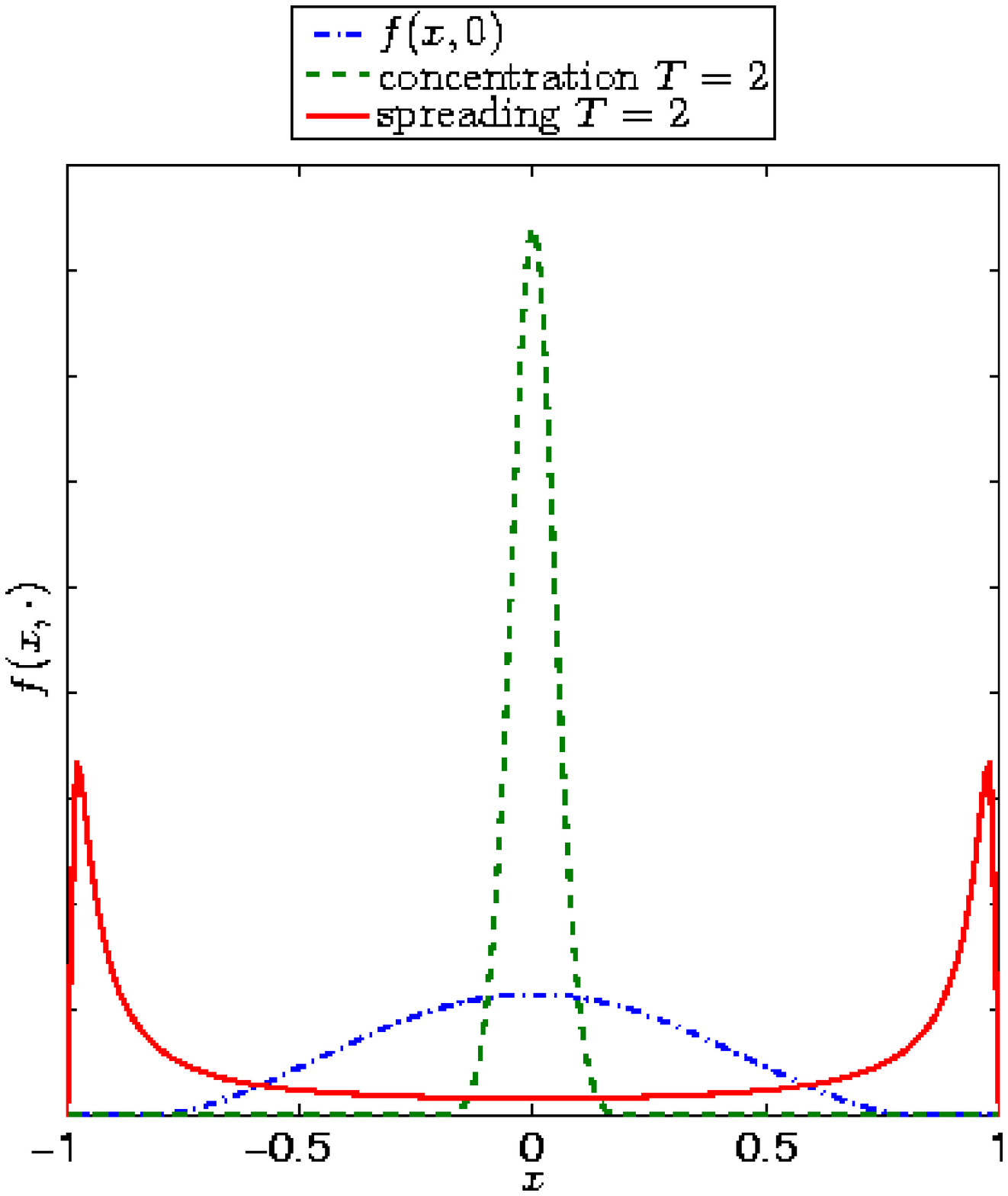}
    \includegraphics[width=.48\textwidth]{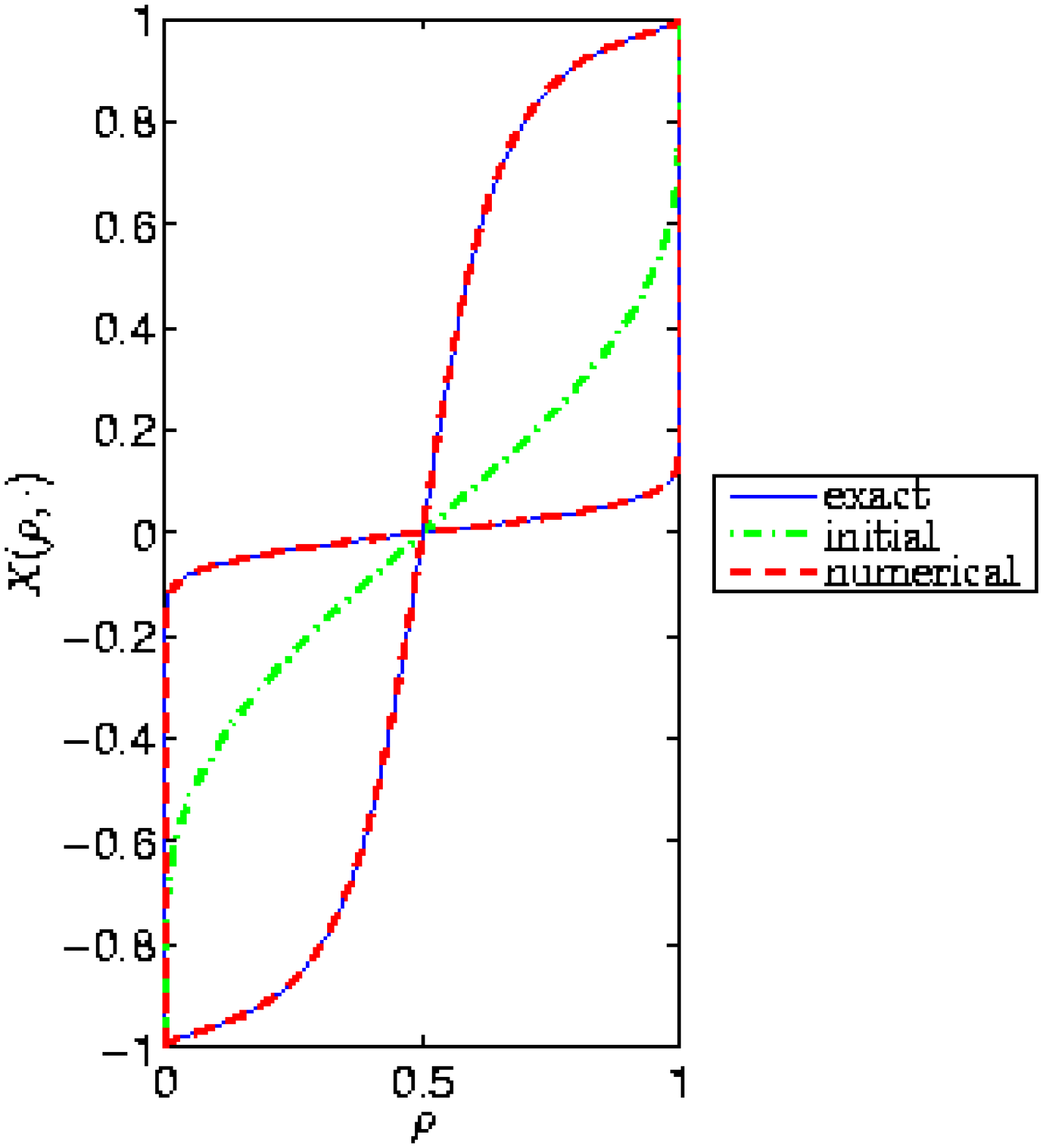}\\
    \caption{Benchmark case: evolution of density function (left hand side figure)
    and comparison between analytical and numerical solution for the quantile function
    (right hand side figure)}
    \label{fig:001}
\end{figure}

 The behavior of \fer{sol5} shows the formation of two peaks in correspondence to
 the extremal points $\pm 1$, while in all other points of the interval $\I$
 there is exponential decay to zero. We remark that equation \fer{magnet-lin}
 has already been solved in \cite{SL}, even if the solution proposed in
 that paper is clearly wrong, as one can easily argue from the fact that it does not
 satisfy the mass conservation property.

 Using the same procedure as above, we can easily solve the problem in
the opposite case of concentration, where $\g= 1$. In this case, if $X_0(\r)$
denote the initial datum,
 \be\label{sol1+}
 X(\r,t) = \frac{X_0(\r) e^{-t}}{\left(1-X_0^2(\r) + X_0^2(\r)e^{-2t}
 \right)^{1/2}}.
 \ee
 The solution now converges exponentially in time to zero, except for
$\r$ values for which $X_0(\r) = \pm
 1$, where it remains constant. In the original formulation, the solution
 $f(x,t)$ corresponding to the initial density   $f_0(x)$ $x \in \I$ is
 \be\label{sol5+}
 f(x,t) =  \frac{e^{-2t}}{\left((1-x^2)e^{-2t} +x^2
 \right)^{3/2}}f_0 \left(\frac{x}{\left((1-x^2)e^{-2t} +x^2
 \right)^{1/2}}  \right).
 \ee
 Note that, except for $x=0$, $f(x,t)$ converges exponentially to zero. If
 $f_0(0) >0$, the solution shows the formation of a peak in $x=0$.

We perform numerical simulation for different initial data in the
general nonlinear case. First, we assume an initial symmetric
datum as a benchmark (see Figure~\ref{fig:001}), where
$$
f_0(x) =
\begin{cases}
c_0 (1-x^2)(0.64-x^2)^{1.3} & \text{if $|x|<0.8$;}\\
0 & \text{otherwise.}\\
\end{cases}
$$
In this case we have $m(t)=0$ for all time $t$. Then we know the
exact solution in order to perform comparison with numerical
results. In Figure~\ref{fig:001} we show the behavior of exact
solution and numerical one. The last one is obtained by using
standard stiff Runge--Kutta methods which is justified by our
theoretical constructive result stated in the proof of
Theorem~\ref{th:exist}. As one can see in Figure~\ref{fig:001} we
have a good agreement between analytical and numerical solution.

\begin{figure}[ht]
    \centering
    \includegraphics[width=.48\textwidth]{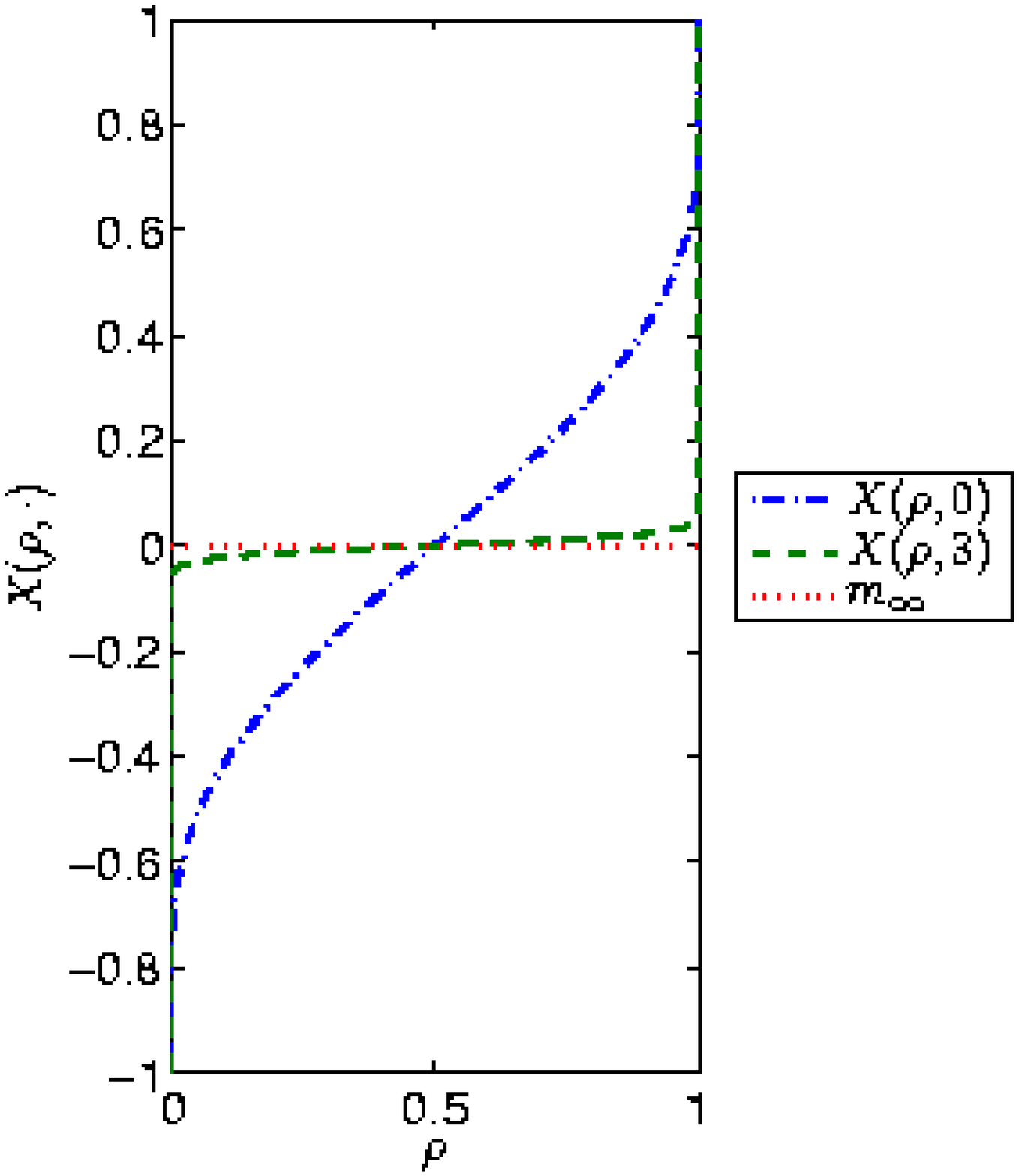}
    \includegraphics[width=.48\textwidth]{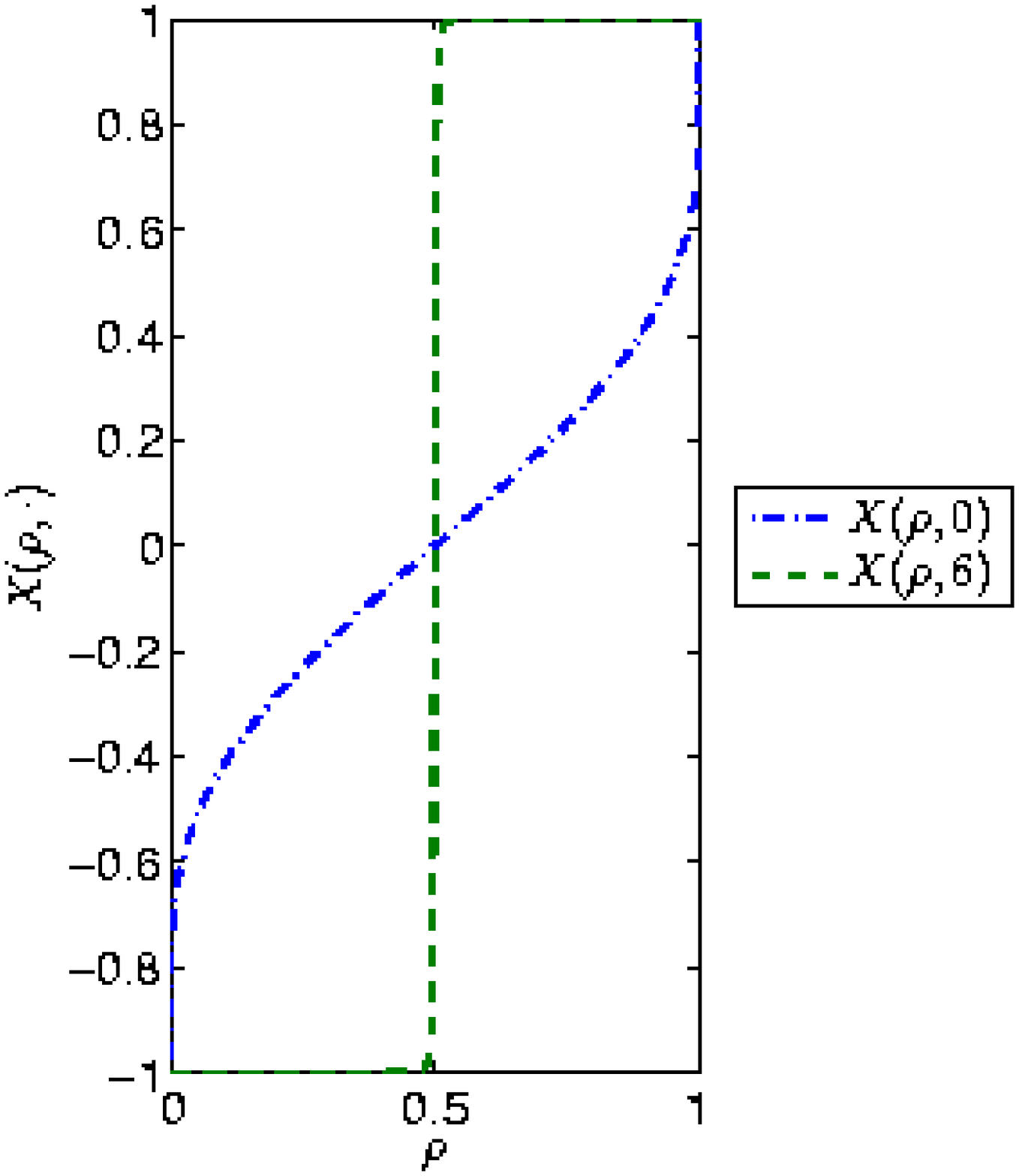}\\
    \caption{Plot of the behavior of numerical solution with symmetric
    initial data for concentration case (left hand side figure), and for
    spreading case (right hand side figure)}
    \label{fig:111}
\end{figure}

In Figure~\ref{fig:111} we sketch the plot of the quantile
function $X(\r,t)$ for different times $t$ both in concentration
and spreading case with the same initial symmetrical data. As
expected, in the concentration case, the limit value of all
quantiles numerically converge to $m_\infty=0$ while in the
spreading case the quantiles converge to $-1$ if $\r<.5$, to $1$
if $\r>.5$.

\begin{figure}[ht]
    \centering
    \includegraphics[width=.48\textwidth]{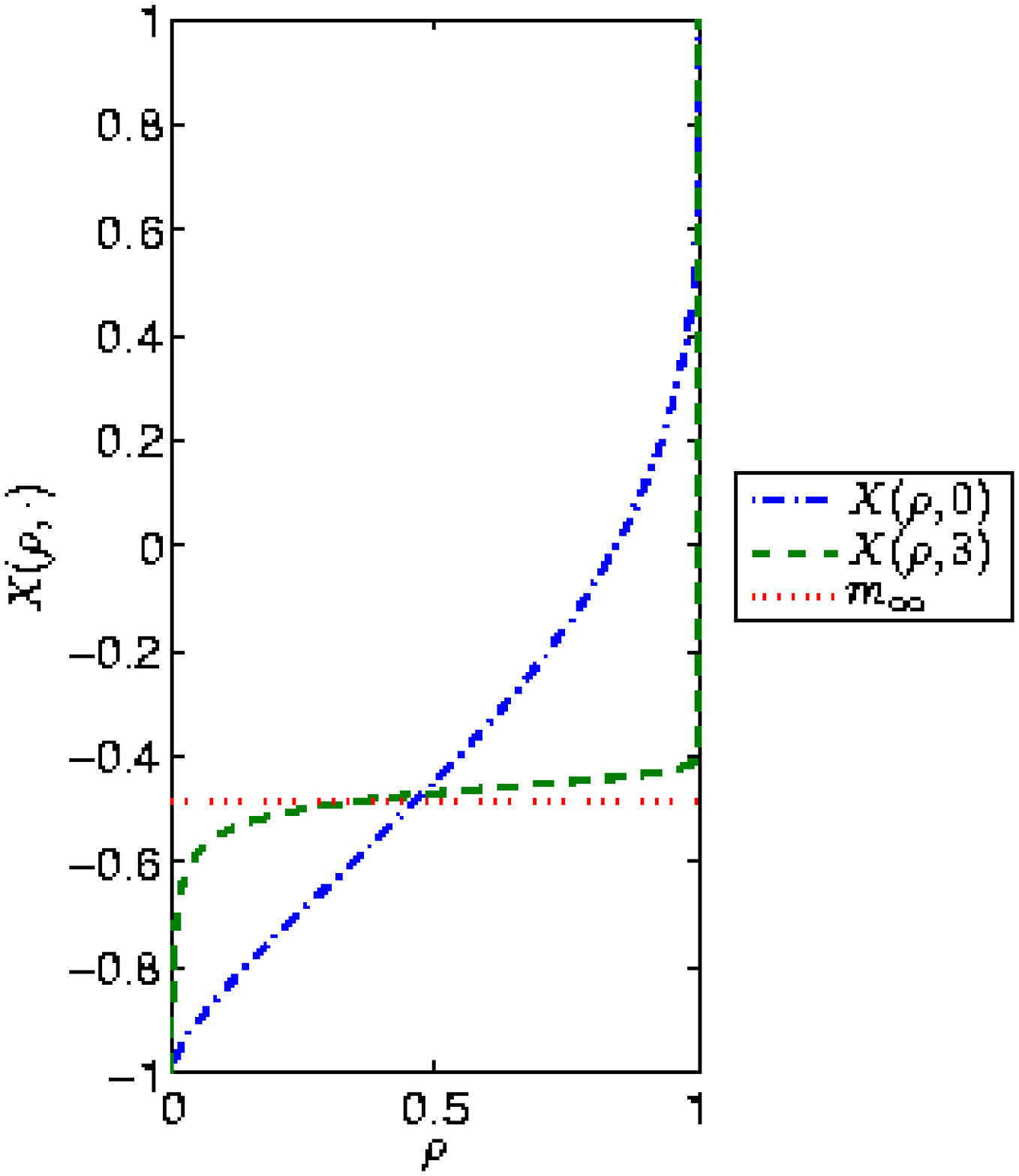}
    \includegraphics[width=.48\textwidth]{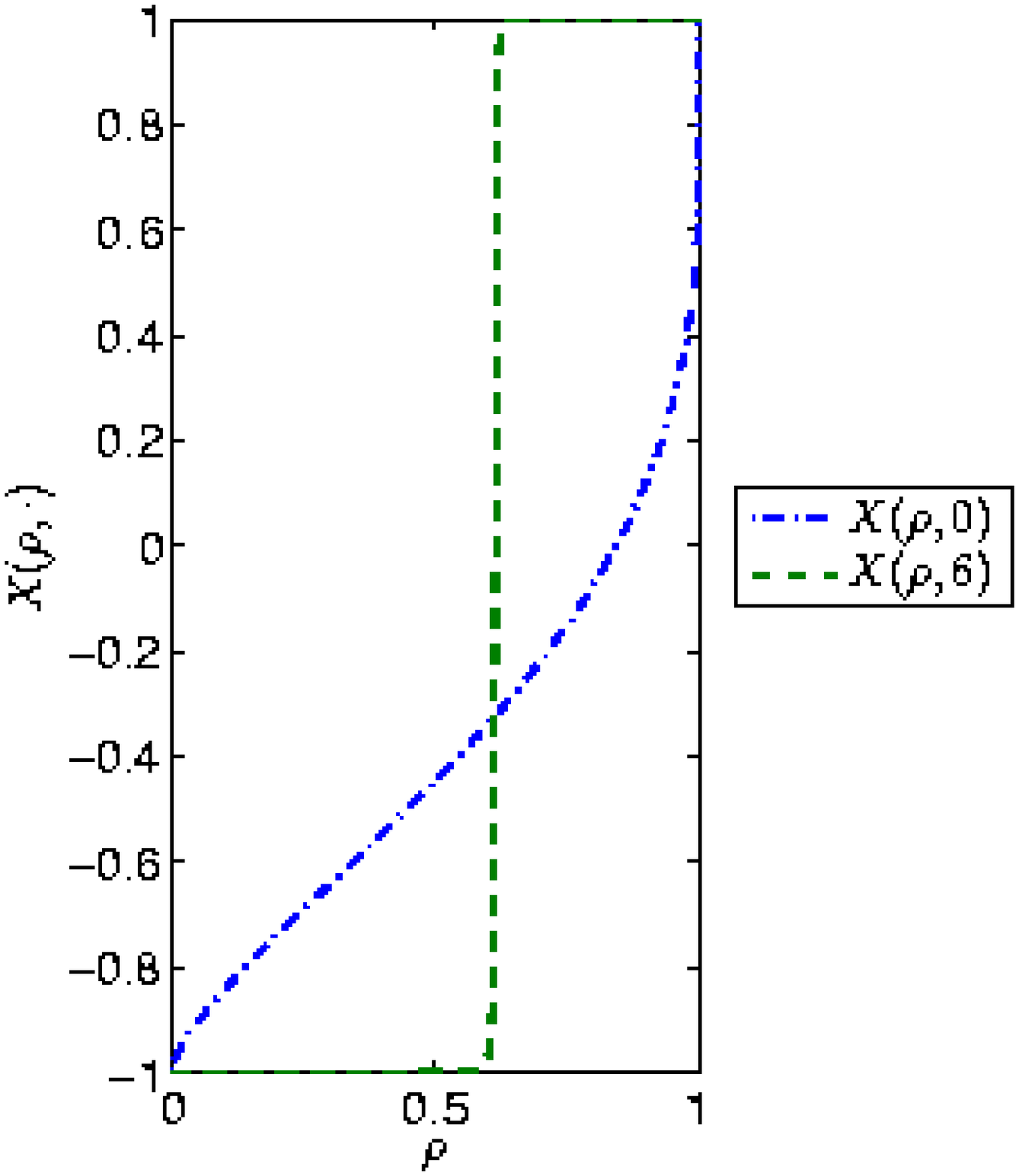}\\
    \caption{Plot of the behavior of numerical solution with non symmetric
    initial data for concentration case (left hand side figure), and for
    spreading case (right hand side figure)}
    \label{fig:222}
\end{figure}

In Figure~\ref{fig:222} we show an asymmetric case. Now, in the
concentration case $m_\infty\neq 0$ given in \eqref{eq:m_infty}
and quantiles converge to it. In the spreading case we can
numerically estimate the value $p^h_{-1}$, see
\fer{eq:monot_lim_X1}, for which $\r<p^h_{-1}$ implies $X(\r,t)\to
-1$, $\r>p^h_{-1}$ implies $X(\r,t)\to 1$.

\begin{figure}[ht]
    \centering
    \includegraphics[width=.48\textwidth]{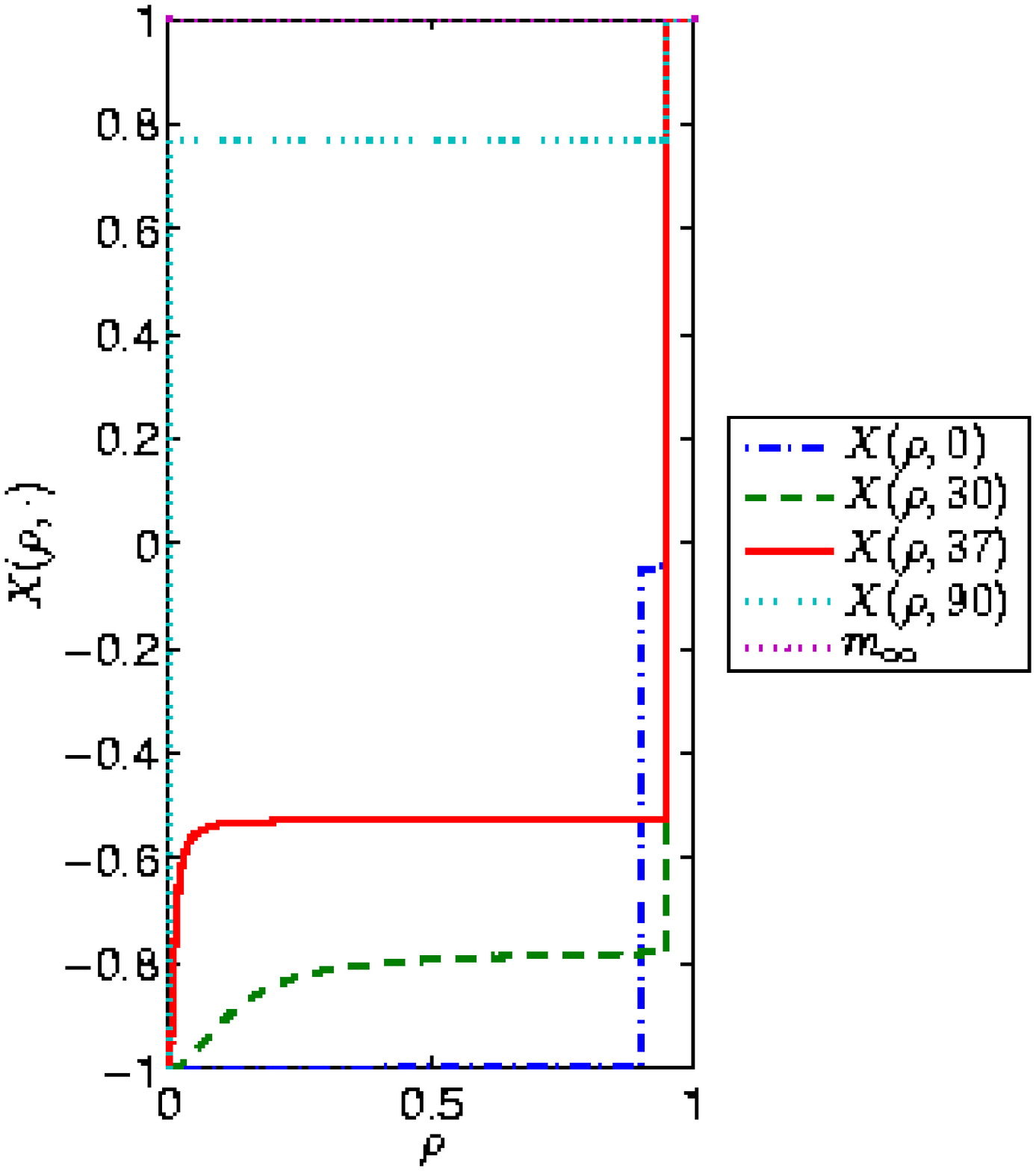}
    \includegraphics[width=.48\textwidth]{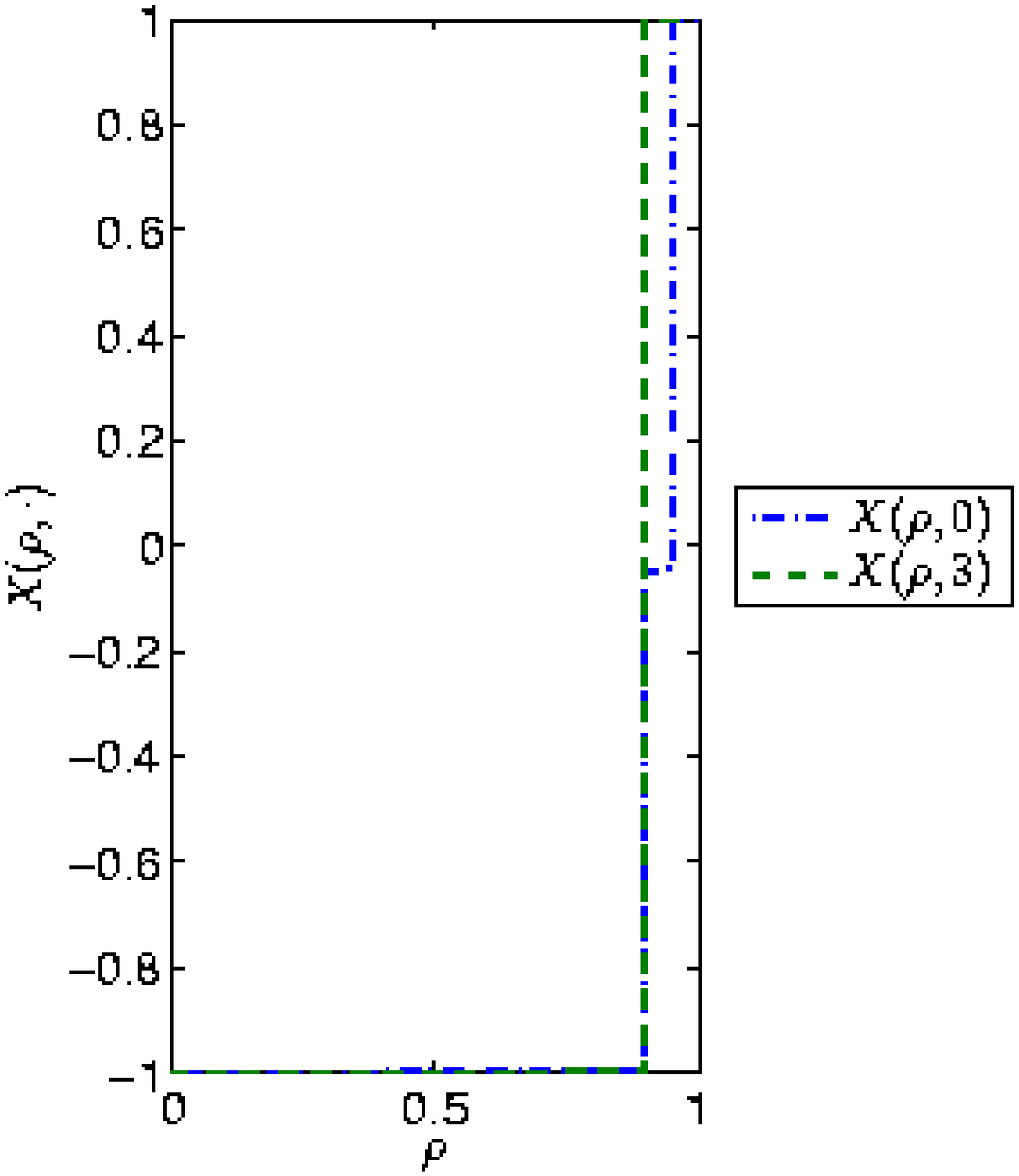}\\
    \caption{Plot of the behavior of a numerical ``controintuitive''
    solution: concentration case tends to $\delta_1$ (left hand side figure),
    while $1-2\epsilon$ mass goes to $-1$ in the
    spreading case (right hand side figure)}
    \label{fig:333}
\end{figure}

Finally, in Figure~\ref{fig:333} we show a numerical
``controintuitive example'' analogous to the example given at the
end of the previous section. Here, the initial datum consists of
$(1-2\epsilon)$ mass very close to $-1$, $\epsilon$ mass close to
$0$, and the remaining $\epsilon$ mass concentrated in $+1$. In
the concentration case, $\gamma=1$, the mass goes asymptotically
to $+1$ while in the spreading case, $\gamma=-1$, only $2\epsilon$
goes asymptotically to $+1$. We point out that the spreading case
is really controintuitive. In fact, if we take $X(\r,T)$, with
$T\gg 1$, solution of the concentration case as initial data for
the spreading one, the dynamic is as follows. Initially,
$(1-\epsilon)$ mass decays, then it splits into two parts: the
right one goes to $+1$, the left one goes to $-1$. The initial
data can be chosen as close as we want to the distribution
$\delta_1$. Then, starting from this data (for which
$T(t)=+\infty$, see \eqref{eq:m_infty}), a $(1-2\epsilon)$ mass
will reach $-1$. We note that the asymptotic state for spreading
phenomena seems unpredictable for such concentrated initial data.

\Remark The method we used to solve equation \fer{magnet-lin}
represents in various case a possible alternative to more known
methods (like the method of characteristics) able to reckon the
solution to one--dimensional first--order partial differential
equations of the form
 \be\label{mag-lin}
 \frac{\partial f}{\partial t} = \frac{\partial
}{\partial x}\left(\phi(x)
 f\right).
 \ee
Our analysis is possible in all cases where the ordinary differential equation
 \[
 \frac{dX}{dt} = -\phi(X)
 \]
 is explicitly solvable.

\section{Conclusions}

We investigated in this paper the spreading and/or  the
concentration of opinion in an organized society by means of a
first--order nonlinear partial differential equation recently
introduced in \cite{to2}. The presence of the nonlinearity render
it difficult to treat analytically the spreading case, and
suitable numerical methods are discussed, able to capture the
large--time behavior of the solution in this case. This work
represents a first attempt for continuous approach to the
formation of opinion in a community of agents. More complete
models can be obtained by considering in addition the (linear or
nonlinear) diffusion, which allows for a continuous steady state
distribution function. Related problems in presence of diffusion
are presently under study.


\end{document}